# Noninvasive magnetic detection of 2D van der Waals room-temperature ferromagnet $Fe_3GaTe_2$ using divacancy spins in SiC

Xia Chen[1,†], Qin-Yue Luo[1,†], Pei-Jie Guo[1], Hao-Jie Zhou[1], Qi-Cheng Hu[1], Hong-Peng Wu[2,3], Xiao-Wen Shen[2,3], Ru-Yue Cui[2,3], Lei Dong[2,3], Tian-Xing Wei[1], Yu Hang Xiao[1], De-Ren Li[1], Li Lei[4], Xi Zhang[1], Jun-Feng Wang[1,*] Gang Xiang[1,*]

[1]College of Physics, Sichuan University, Chengdu 610065, China

[2]State Key Laboratory of Quantum Optics and Quantum Optics Devices, Institute of Laser Spectroscopy, Shanxi University, Taiyuan 030006, China

[3]Collaborative Innovation Center of Extreme Optics, Shanxi University, Taiyuan 030006, China

[4]Institute of Atomic and Molecular Physics, Sichuan University, Chengdu 610065, China

[†]These authors contributed equally: Xia Chen, Qin-Yue Luo.

[*]Corresponding author: jfwang@scu.edu.cn; gxiang@scu.edu,cn

**Abstract**

Room-temperature (RT) two-dimensional (2D) van der Waals (vdW) ferromagnets hold immense promise for next-generation spintronic devices for information storage and processing. To achieve high-density energy-efficient spintronic devices, it is essential to understand local magnetic properties of RT 2D vdW magnets. In this work, we realize noninvasive *in situ* magnetic detection in vdW-layered ferromagnet $Fe_3GaTe_2$ using divacancy spins quantum sensor in silicon carbide (SiC) at RT. The structural features and magnetic properties of the $Fe_3GaTe_2$ are characterized utilizing Raman spectrum, magnetization and magneto-transport measurements. Further detailed analysis of temperature- and magnetic field-dependent optically detected magnetic resonances of the PL6 divacancy near the $Fe_3GaTe_2$ reveal that, the Curie temperature ($T_c$) of $Fe_3GaTe_2$ is ~360 K, and the magnetization increases with external magnetic fields. Additionally, spin relaxometry technology is employed to probe the magnetic fluctuations of $Fe_3GaTe_2$, revealing a peak in the spin relaxation rate around the $T_c$. These experiments give insights into the intriguing local magnetic properties of 2D vdW RT ferromagnet $Fe_3GaTe_2$ and pave the way for the application of SiC quantum sensors in noninvasive *in situ* magnetic detection of related 2D vdW magnets.

**Introduction**

In recent years, numerous innovative magnetic materials such as van der Waals (vdW) magnets[1], topological insulators[2] and high-temperature superconductors[3] have been discovered and drawn extensive attention[1-5]. Among those, various vdW ferromagnetic materials have emerged as promising platforms for two-dimensional (2D) high-density low-power spintronic devices due to their remarkable properties, including high degrees of freedom encompassing spin, charge and intrinsic magnetism[6-8]. However, a common limitation faced by most 2D vdW ferromagnets lies in their low Curie temperatures ($T_c$), which restricts their practical applications in spintronic devices[6-10]. Recently, the discovery of room-temperature (RT) vdW ferromagnet $Fe_3GaTe_2$ has captured significant interest due to its robust intrinsic ferromagnetism[9,10].

The growing novel magnetic materials inspire scientists to develop new technologies to investigate the local magnetic fields generated by spins and electric currents in these materials[4,5,11]. Traditional magnetic field detection technologies, such as nuclear magnetic resonance (NMR) and magnetic force microscopy (MFM), have played crucial roles in studying magnetic materials. However, these methods suffer from limitations, including poor spatial resolution, narrow operating temperature ranges, and perturbative effects from magnetic probes[4,5]. To address these challenges, some spin qubits in solid-state including nitrogen-vacancy (NV) centers in diamond and boron vacancy ($V_B^-$) defects in hexagonal boron nitride (hBN) have been exploited as versatile magnetic quantum sensors[4,5,12-14]. Owing to the advantages of noninvasive, high sensitivity, nanoscale spatial resolution and wide operating temperature range, spin qubits have been applied for detecting static and dynamic magnetic field in various materials, encompassing ferromagnetic materials, superconductors, and topological insulators, even under high pressure[4,5,12-15]. However, integrating these spin qubits with practical electronic devices is still of great challenge. The diamond and hBN substrates lack mature growth and micro/nano fabrication process as well as compatibility with complementary metal-oxide-semiconductor (CMOS) technology. Fortunately, recent accomplishments have highlighted silicon carbide (SiC) as a promising solid-state system for quantum technologies[16-22]. SiC is a mature semiconductor, characterized by well-established inch scale single-crystalline growth and subsequent doping and device fabrication protocols[16-22]. Moreover, SiC enjoys widespread adoption in high-power electronic devices, making it an ideal candidate for integration with 2D vdW ferromagnetic materials in spintronic and electronic applications.

The spin qubits in SiC, including silicon vacancy, divacancy, NV centers and transition-metal color centers[16-23], have been exploited in quantum photonics, quantum information processing, quantum network and quantum sensing due to their near infrared fluorescence and remarkable coherence properties, even at RT[19-23]. In the realm of quantum sensing, these spin qubits serve as robust and versatile nanoscale sensors, capable of measuring various physical quantities such as magnetic fields, electric fields, strain, pressure, and temperature[19-22,24-29]. For instance, nanotesla magnetometry have been realized using optically detected magnetic resonance (ODMR) technologies of the silicon vacancy ensemble[24]. Additionally, SiC-based anvil cells enable the *in situ* magnetic detections for superconductors under high pressure[25]. The charge states and ODMR approaches of divacancies have been applied to realize electric field sensing[26,27]. High-sensitivity temperature sensing (14 mK/Hz$^{1/2}$) is also realized through the utilization of thermal dynamical decoupling methods[28]. Recently, divacancies in SiC have exhibited wide-range pressure sensing capabilities with a sensitivity of 0.28 MPa/Hz$^{1/2}$ at RT[29]. Despite these advances, most quantum sensing concentrates on probing various physical quantities, while little is applied to probe magnetic properties in condensed matter physics[19-22,24-29]. By leveraging SiC's mature technologies, we can expand the scope of SiC-based quantum sensing and accelerate the development of vdW ferromagnet-based spintronic and electronic devices.

In this work, we achieve the noninvasive *in situ* local magnetic detection of the 2D vdW-layered $Fe_3GaTe_2$ using divacancy spins in SiC at RT, where a layer of shallow divacancies serves as the local magnetic probes of small pieces of $Fe_3GaTe_2$ flakes. The lattice vibration, magnetization and anomalous Hall resistance properties of the $Fe_3GaTe_2$ sample are firstly characterized. We then compare the temperature-dependent ODMR signals from divacancies located near and far from the $Fe_3GaTe_2$ sample, and reveal that the $T_c$ of $Fe_3GaTe_2$ is ~360 K. Additionally, the magnetic field-dependent ODMR signals are measured to study the evolution of magnetization with magnetic field. Finally, a peak in the spin relaxation rate around the $T_c$ is discovered through the temperature-dependent spin relaxometry. These experiments lay the groundwork for the applications of mainstream semiconductor technology-friendly SiC-based quantum sensors to noninvasive *in situ* local magnetic detection of 2D vdW ferromagnets.

**Results**

The lattice vibration, elemental composition and stoichiometry measurements are first performed to identify the $Fe_3GaTe_2$ sample. Fig. 1a. illustrates the crystalline structure of $Fe_3GaTe_2$. The vdW-layered $Fe_3GaTe_2$ presents a hexagonal structure ($a=b=3.99$ Å, $c=16.23$ Å, $α=β=90°$, $γ=120°$) with $p6_3/mmc$ space group, where each $Fe_3Ga$ layer is encapsulated by two Te layers and the adjacent 0.78-nm-thick Te layers are bonded by weak vdW force[30]. The Raman spectroscopy of the vdW-layered $Fe_3GaTe_2$ in Fig. 1c. shows two prominent Raman peaks at 125.8 cm$^{-1}$ and 143.0 cm$^{-1}$ under a 25 mW excitation laser, corresponding to the $A_{1g}$ and $E_{2g}$ vibrational modes of $Fe_3GaTe_2$[31], respectively. The absence of extraneous peaks corroborates the high crystal quality of the $Fe_3GaTe_2$ sample. The scanning electron microscope (SEM) picture and the correlated energy dispersive X-ray spectroscopy (EDS) mapping in Fig. 1d. indicate that, the flattened $Fe_3GaTe_2$ sample exhibits a uniform elemental distribution and a moderate atomic ratio of Fe, Ga and Te elements is 3.14:1.00:1.97, well consistent with the stoichiometric ratio.

Then, the intriguing RT ferromagnetic properties are examined by a magnetic properties measurement system (MPMS, Quantum Design). Fig. 1e. presents the out-of-plane (OP) and in-plane (IP) temperature-dependent magnetization curves under zero-field cooling (ZFC) and field-cooling (FC) conditions, exhibiting a typical magnetic phase transition from paramagnetism to ferromagnetism at the $T_c$ of ~360 K, consistent with the reported $T_c$ of single-crystalline $Fe_3GaTe_2$ samples[9,31]. An apparent preponderance on OP magnetization than IP one indicates the perpendicular magnetic anisotropy (PMA) in the $Fe_3GaTe_2$. The RT ferromagnetism and PMA in the $Fe_3GaTe_2$ are also explicitly implied by magnetic field-dependent OP and IP magnetization (*M-H*) curves at 300 K, as shown in Fig. 1f. Meanwhile, the magnetic domain structures are characterized by magnetic force microscopy (MFM) at RT in the absence of external magnetic field (details can be seen in Supplementary S1), which reveals the occurrence of a single- (multi-) domain state in thin (thick) $Fe_3GaTe_2$ flakes due to the competition among dipole interaction, exchange interaction and magnetic anisotropy[31], underscoring the robust RT ferromagnetism of the $Fe_3GaTe_2$.

Since the well-established RT PMA of the $Fe_3GaTe_2$ can also be confirmed by magneto-transport properties, next, longitudinal resistivity ($ρ_{xx}$) and transverse resistivity ($ρ_{xy}$) of the $Fe_3GaTe_2$-based Hall bar device are conducted (fabrication details can be seen in Supplementary S2). Figs. 1g, h, present the temperature-dependent $ρ_{xx}$, where the $ρ_{xx}$ decreases with decreasing temperature, revealing the metal conducting behavior of the $Fe_3GaTe_2$ nanoflake. Interestingly, the inflection point of the $ρ_{xx}$ occurs at ~20 K, indicating the presence of Kondo scattering[31,32]. Additionally, the OP magnetic field-dependent $\rho_{xy}$ at the temperatures ranging from 5 K to 360 K in Fig. 1i. reveals near-square hysteresis loops, where discontinuous magnetization switching may originate from the multi-domain magnetic structures[33]. These characterization results again indicate the large PMA and high $T_c$ of Fe$_3$GaTe$_2$. The above-mentioned structural, magnetic or electronic transport measurements reveal the good crystalline quality, robust RT ferromagnetism and large PMA of the vdW-layered $Fe_3GaTe_2$, making it an ideal RT ferromagnet for low-dimensional spintronic and electronic devices.

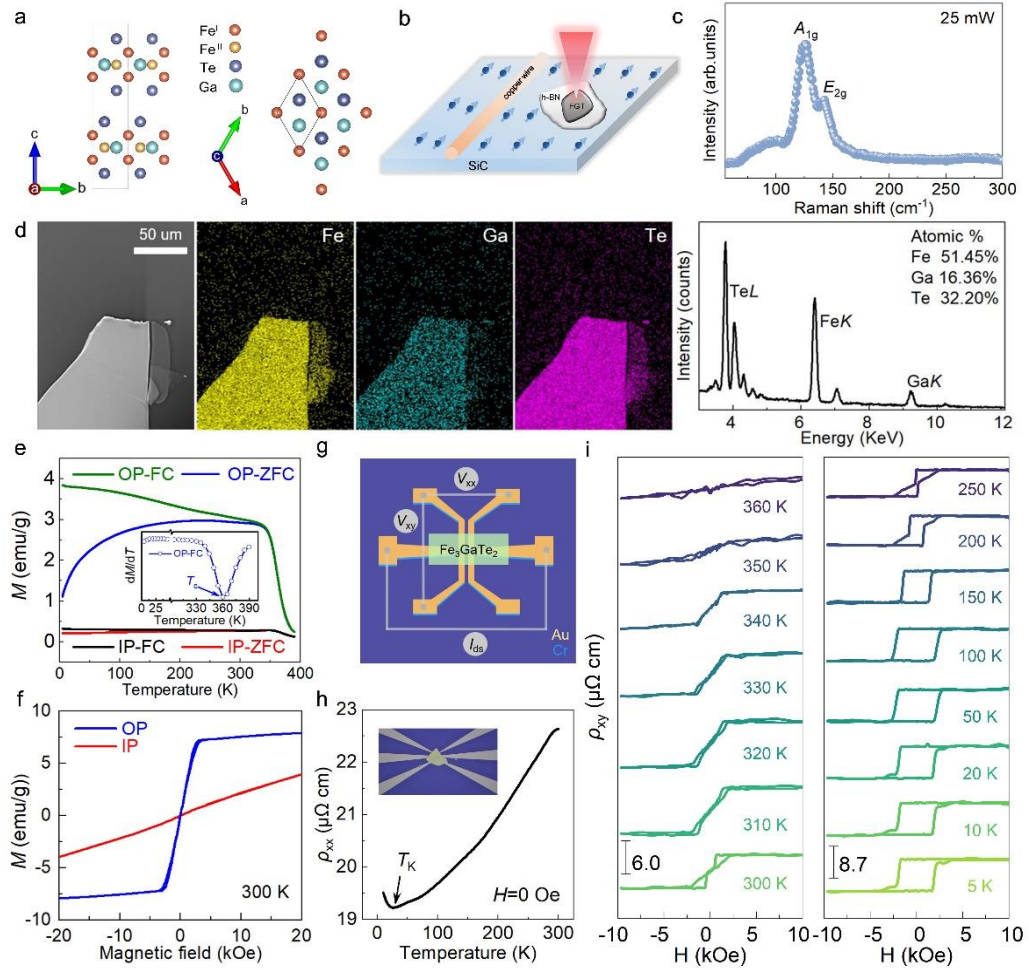

**Fig. 1 Characterizations of RT ferromagnet vdW-layered Fe$_3$GaTe$_2$.** **a** Crystal structure of Fe$_3$GaTe$_2$ from *bc*-plane and *ab*-plane views. **b** Schematic of an hBN/Fe$_3$GaTe$_2$ heterostructure transferred onto SiC, where shallow divacancy defects are used for magnetic detection. **c** Raman spectrum of Fe$_3$GaTe$_2$. **d** SEM image of an exfoliated Fe$_3$GaTe$_2$ flake on SiO$_2$/Si substrate with its corresponding EDS spectrum and EDS mapping of Fe, Ga and Te atoms. **e** Temperature-dependent OP and IP magnetization curves of Fe$_3$GaTe$_2$ under ZFC and FC (1000 Oe) conditions, and the inset shows the obtained $T_c$ using temperature-dependent d$M$/d$T$ curve under OP-FC condition. **f** At 300 K, OP and IP *M-H* curves of Fe$_3$GaTe$_2$. **g** Schematic diagram of the Fe$_3$GaTe$_2$ Hall bar device. **h** Temperature-dependent longitudinal resistivity of the Fe$_3$GaTe$_2$ Hall bar device, where $T_K$ represents the Kondo temperature. **i** Temperature-dependent anomalous Hall resistivity of the Fe$_3$GaTe$_2$ Hall bar device under OP magnetic field.

After characterizing the basic physical properties of the $Fe_3GaTe_2$, we then perform the noninvasive local magnetic detection of it using the PL6 divacancies in 4H-SiC. There are seven types of divacancies (PL1 to PL7) in 4H-SiC, and their spin state S is 1[34]. The zero-phonon line (ZPL) and zero field splitting (ZFS) D value of PL6 divacancy spins are 1038 nm[28] and 1351 MHz[35], respectively. Its direction is along the *c*-axis of the 4H-SiC. The schematic of SiC magnetic detection of the $Fe_3GaTe_2$ sample is presented in Fig. 1b. A small flake of $Fe_3GaTe_2$ is placed on the surface of the SiC, where a shallow layer of PL6 divacancy spins is used to detect the $Fe_3GaTe_2$ sample. A hBN flake is placed on $Fe_3GaTe_2$ to protect it from oxidation. First, the temperature-dependent magnetization of the $Fe_3GaTe_2$ is investigated using the ODMR methods. Figs. 2a. and 2b. show the optical microscopy image and confocal scanning image of the $Fe_3GaTe_2$ sample, respectively. As shown in Fig. 2b, the probe position (red cross) is close to $Fe_3GaTe_2$ sample and the reference position (black cross) is far away from the sample.

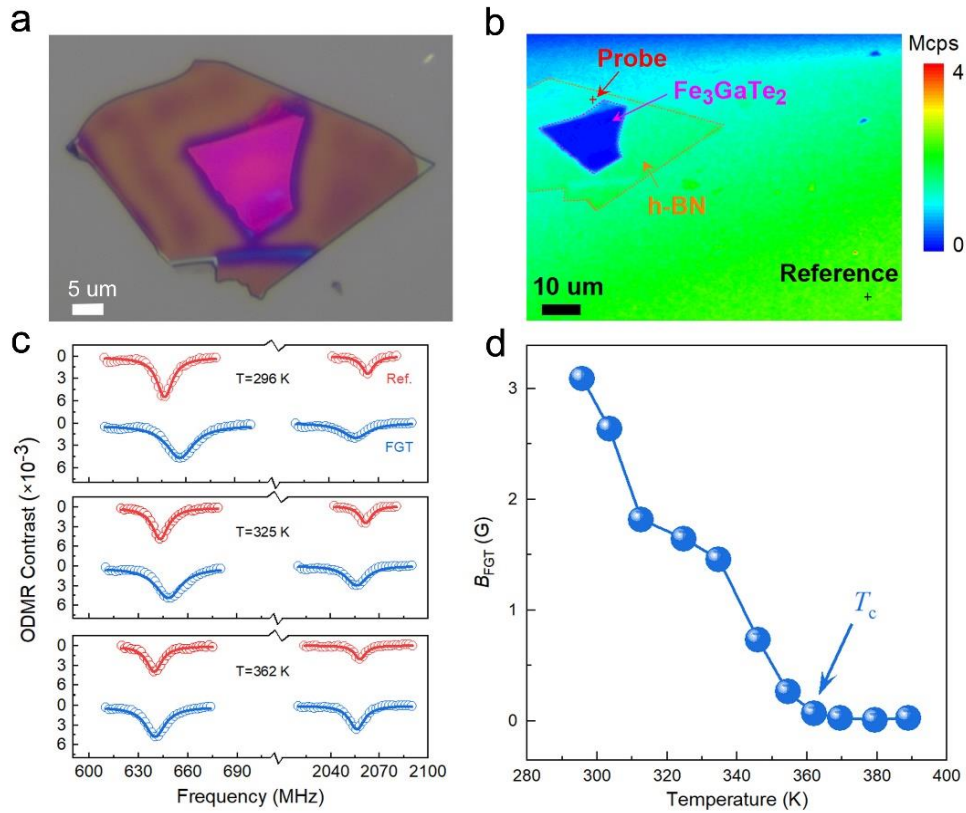

**Fig. 2 Magnetic detection of the $Fe_3GaTe_2$ with temperature. a** Image of the $Fe_3GaTe_2$ sample under an optical microscope. The large brown substance covering the $Fe_3GaTe_2$ (pink sample) is hBN. **b** Confocal scanning microscopy image of the $Fe_3GaTe_2$ sample. Orange lines outline the boundary of hBN, within which the deep blue area represents the $Fe_3GaTe_2$ sample, in agreement with optical image. The red and black crosses near and far away from $Fe_3GaTe_2$ represent the probe position and reference position, respectively. **c** ODMR measurements at the probe position and at the reference position under different temperatures and a magnetic field of ~200 G. Hollow dots are experimental data and the solid lines are the corresponding Lorentz fits. **d** The magnetic fields of $Fe_3GaTe_2$ $B_{FGT}$ as a function of temperature. From the experiments, we obtain the $T_c$ is ~360 K.

We perform the ODMR measurements as a function of temperature with a *c*-axis external magnetic field $B_0$ of ~200 G. For clarity, we define the $B_{tot}$ and $B_{FGT}$ as the total and $Fe_3GaTe_2$ magnetic fields, respectively. Inferred from the ODMR splitting at the probe and reference positions, we can obtain the corresponding $B_{tot}$ and $B_0$, respectively. Then the magnetic field of $Fe_3GaTe_2$ sample $B_{FGT}$ is $|B_{tot} - B_0|$[12,13,25,36]. Fig. 2c. shows three pairs of ODMR results at the probe position and the reference position under different temperatures. The difference values between the probe and reference positions obviously decrease with increasing temperature. In addition, all the ODMR peak values decrease as the temperature increases due to the decrease of the ZFS D[37]. The inferred $B_{FGT}$ with respect to temperature is displayed in Fig. 2d. The $B_{FGT}$ decreases from about 3.2 G to about 0 G as the temperature increases from 295.8 K to about 360 K, and then remains at about 0 G as the temperature continues to increase to about 389 K. Inferred from the experiments, the $T_c$ of $Fe_3GaTe_2$ is ~360 K, which is consistent with previous results[9].

In addition to temperature, the external magnetic field also affects the magnetism of the 2D vdW $Fe_3GaTe_2$[9,10]. We then perform ODMR measurements under different external magnetic fields at RT to investigate the relationship between $B_{FGT}$ and the external magnetic field. The confocal scanning microscopic image of the PL6 and $Fe_3GaTe_2$ is shown in Fig. 3a. We then study the ODMR at the probe position near the $Fe_3GaTe_2$ (blue cross) and the reference position (black cross). Fig. 3b. presents three pairs of the ODMR measurement results. Both the difference values of the ODMR left and right branches between the reference and probe positions increase as the external magnetic field increases, which means that the $B_{FGT}$ at the probe position increases with increasing external magnetic field. At the same time, we also measure the ODMR under different negative magnetic fields. Fig. 3c. summarizes the $B_{FGT}$ as a function of external magnetic field, and $B_{FGT}$ increases with increasing positive and negative magnetic field. Inferred from the experiments, we obtain the coercive field of this sample to be ~10 G, which is consistent with the *M-H* curve in Fig. 1f.

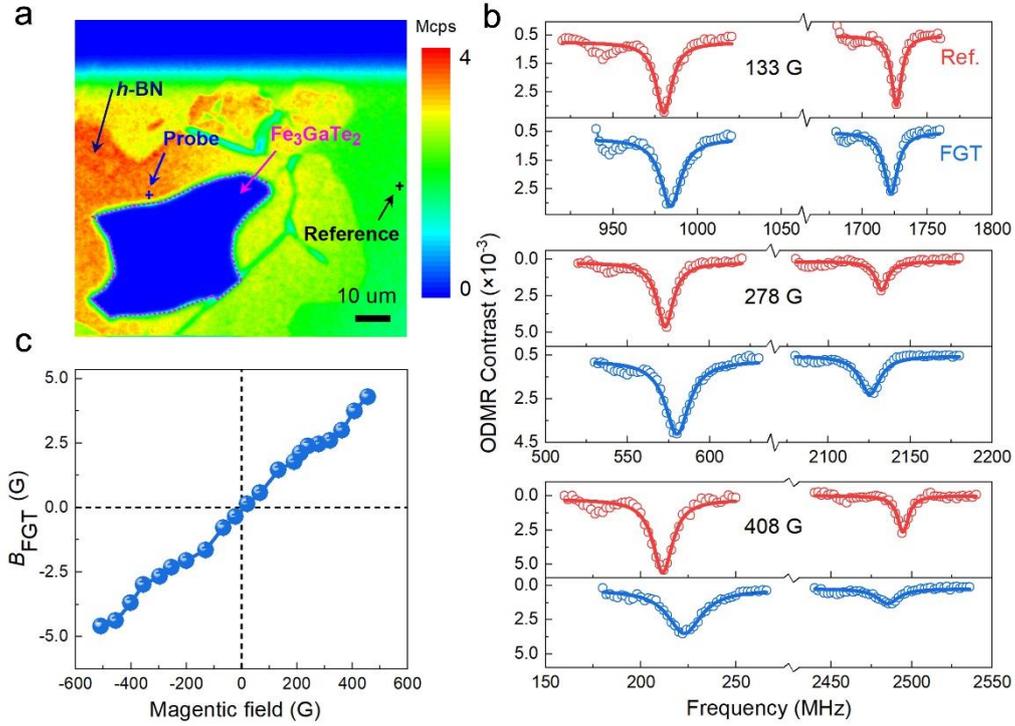

**Fig. 3 Magnetic detection of the Fe$_3$GaTe$_2$ sample with external magnetic field. a** Confocal scanning microscopy image of SiC surface, where the blue and black crosses represent the probe position and the reference position, the deep blue polygon is the Fe$_3$GaTe$_2$ sample and the rectangular dark blue area at the top of the image is the copper wire. **b** ODMR measurements at the probe position and the reference position under three representative external magnetic fields at RT. The hollow dots are the experimental data and the solid lines are Lorentz fits. **c** $B_{FGT}$ at the probe position as a function of external magnetic fields ranging from -508 G to 455 G.

Except for the static magnetic field, the vdW ferromagnet Fe$_3$GaTe$_2$ also exhibits intrinsic dynamic fluctuating magnetic field[4,5,9,10,12,13]. Unlike traditional magnetic detection methods, as a versatile magnetic quantum sensor, spin defects in SiC could detect both the d.c. and a.c. magnetic fields[4,5,9,10,12,13,36]. To understand the dynamical behavior of magnetic spin fluctuations generated by static longitudinal magnetic susceptibility and diffusive spin transport of the Fe$_3$GaTe$_2$[12,13,36], we study the spin relaxometry of PL6 divacancy spins close to the Fe$_3$GaTe$_2$ at different temperatures. The upper panel of Fig. 4a. shows the pulse sequence of the divacancy relaxation measurements[12,13,16]. Three representative PL6 relaxometry results at the reference position under an external magnetic field of ~190 G are displayed in the lower part of

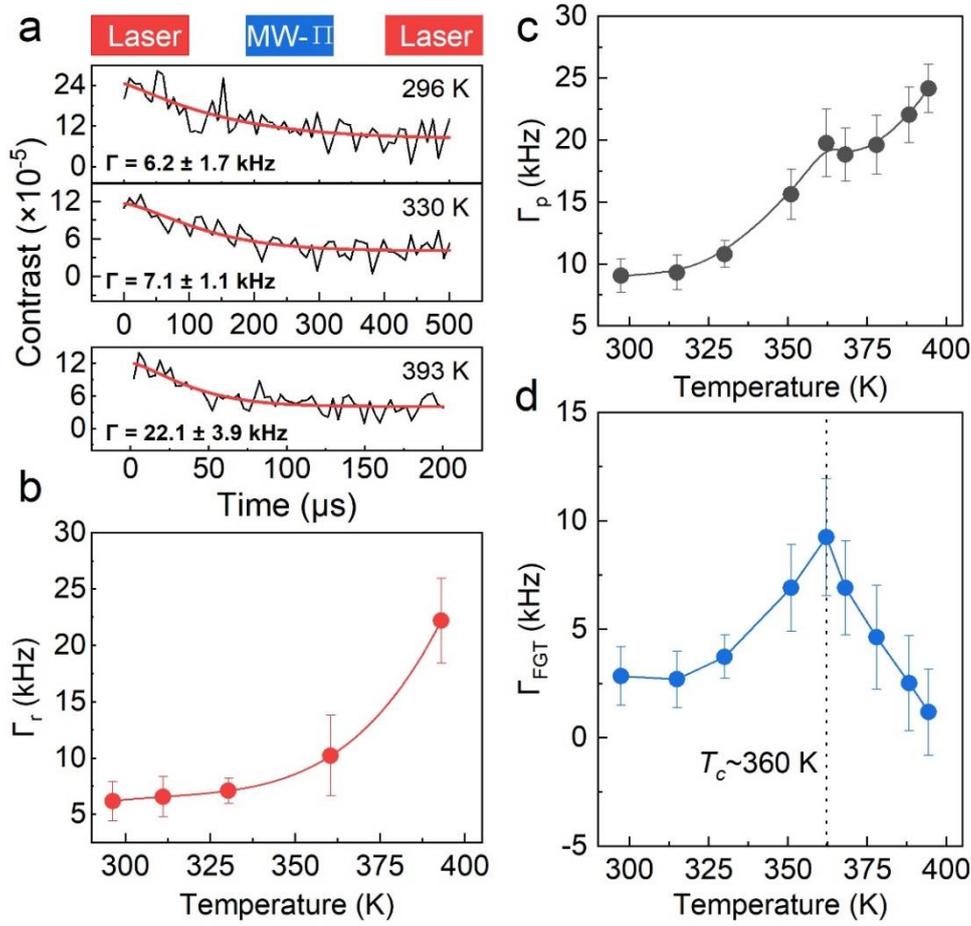

**Fig. 4 Fe$_3$GaTe$_2$ magnetic spin fluctuation detection using spin relaxometry of PL6. a** The spin relaxometry pulse sequence and three measurement results at the reference position under different temperatures. As the temperature increases from 296 K to 393 K, the relaxation rate increases from 6.2 kHz to 22.1 kHz. The black lines are experimental data and the red solid lines are the fits to the data. **b**, **c** Summarization of the relaxation rate from RT to 393K at the reference position **(b)** and at the probe position **(c)**. The red solid line is the fit. **d** Temperature-dependent relaxation rate of the Fe$_3$GaTe$_2$ magnetism, which exhibits a peak around $T_c$.

Fig. 4a, respectively. We use an exponential decay function $e^{-(t\Gamma)^n}$ to fit the data, where $\Gamma$ is the relaxation rate and n is the exponential parameter[12,13,16]. The relaxation rate at RT (296 K) is 6.2 ± 1.7 kHz, which is in agreement with previous results[16]. We denote the PL6 relaxation rates at the probe and reference positions (see Supplementary Information Figure 3 for details) as $\Gamma_p$ and $\Gamma_r$, and the derived Fe$_3$GaTe$_2$ sample $\Gamma_{FGT}$ is $\Gamma_p$-$\Gamma_r$[12,13,16]. It can be seen that the intrinsic relaxation rate $\Gamma_r$ is larger at higher temperature. Fig. 4b. summarizes the intrinsic relaxation rate $\Gamma_r$ at the reference position as a function of temperature. It is obvious that $\Gamma_r$ accelerates its increase with increasing temperatures, which is consistent with previous results of NV center in diamond[38].

Given the temperature-dependent effect of higher-energy lattice phonons two-phonon Raman process and the local phonons two-phonon Orbach-type process in SiC[38], we fit the data with formular: $\boldsymbol{\Gamma_r = a + b/(e^{\Delta/kT} - 1) + cT^5}$, where a, b and c are fitting parameters and a is related to PL6 and its spin bath concentration. The k is Boltzmann's constant, T is temperature and $\Delta$ is the dominant local vibrational energy[38].

As a comparison, we then detect PL6 relaxation rate $\Gamma_p$ at the probe position under different temperatures. Fig. 4c. exhibits the measured $\Gamma_p$ at different temperatures. The $\Gamma_p$ increases faster as the temperature is closer to the $T_c$ of ~360 K, and increases slowly as the temperature further increases. By subtracting the intrinsic spin relaxation rate $\Gamma_r$ at the corresponding temperatures, we can derive the $\Gamma_{FGT}$. Fig. 4d. summarizes the temperature-dependent $\Gamma_{FGT}$. The magnetic fluctuation rate $\Gamma_{FGT}$ increases as the temperature increases close to the $T_c$ of ~360 K, due to the increase of the magnetic susceptibility around the $T_c$ [9,10,12,13]. However, the $\Gamma_{FGT}$ decreases as the temperature further increases to about 393 K, which is due to the fact that the remaining active magnetic susceptibility decreases and the magnetic fluctuations in $Fe_3GaTe_2$ are small when temperature exceeds the $T_c$ [9,10,12,13]. The phenomenon that temperature-dependent $\Gamma_{FGT}$ reveals a peak around the $T_c$ is similar to the other vdW ferromagnets and moiré magnetism[12,13].

**Discussion**

In conclusion, we realize the noninvasive static and dynamic fluctuating magnetic detection of 2D vdW RT ferromagnet $Fe_3GaTe_2$ using PL6 divacancy spins in SiC. Through the temperature- and magnetic field-dependent ODMR measurements near the $Fe_3GaTe_2$ sample, results show that the $T_c$ is ~360 K and the magnetic susceptibility increases with the increasing external magnetic field. Employing spin relaxometry technology, the magnetic spin fluctuations of $Fe_3GaTe_2$ exhibit a peak around the $T_c$ of ~360 K. The versatile PL6 quantum sensor could also be applied to detect the local magnetism of other novel magnetic materials, such as antiferromagnetic materials, superconductors and topological insulators. Combined with infrared CMOS camera[5,12,13] and SiC anvil cell[24,29,36,39], it could realize the magnetic image and magnetic detection of various novel magnetic materials under high pressure. The experiments highlight the significant potential of well-established technologies of SiC spin defects for noninvasive *in situ* investigation of local static and dynamic magnetic behavior in vdW ferromagnets, which could accelerate the development of vdW ferromagnet-based spintronic and electronic devices.

**Methods**

**Sample and device fabrication**

The intrinsic vdW-layered $Fe_3GaTe_2$ crystal is purchased from Nanjing Mknano Tech. Co., Ltd. For the $Fe_3GaTe_2$ Hall bar device, pre-patterned Cr/Au (8 nm/50 nm) electrode with a standard Hall bar geometry is fabricated on a $SiO_2$/Si substrate using lithography technology, thermal evaporation and lift-off technology (see Supplementary Information Figure 2 for details). Therewith, the $Fe_3GaTe_2$ nanoflakes are mechanically exfoliated with the aid of scotch tape from the bulk counterpart and transferred onto the SiC or $SiO_2$/Si substrate by polydimethylsiloxane (PDMS) using transfer system, which are performed in the glove box. To avoid the oxidation of $Fe_3GaTe_2$ sample, a hBN flake is transferred onto the $Fe_3GaTe_2$ flake to cover and protect.

**Physical properties characterizations**

Atomic force microscopy and MFM (Bruker) are used to characterize the thickness and magnetic domain structures of the vdW-layered $Fe_3GaTe_2$. The vibration modes are confirmed by Raman spectroscopy test platform (532 nm solid-state laser, 25 mW excitation Laser). The composition and stoichiometric ratio are obtained by SEM equipped with EDS (JSM-IT500HR). The magnetic properties under various temperatures and magnetic field orientations are performed by superconducting quantum interference device (SQUID, Quantum Design). The $\rho_{xx}$ and $\rho_{xy}$ at different temperatures are detected by an electronic transport measurement system (Model ET9007 from East Changing Technologies, China). During the measurement of $\rho_{xy}$, the magnetic field is applied perpendicular to the sample and the temperatures ranging from 5 K to ~360 K.

**SiC sample and the experimental set-up**

The SiC sample is a high-purity 4H-SiC epitaxial layer sample[23]. We use the nitrogen ion implantation (energy 30 keV, dose $1\times10^{14}$ $cm^{-2}$) to generate a shallow layer of divacancy ensemble (about 40 nm, stopping and range of ions in matter (SRIM)). Then the sample is annealed at 1050 °C for 2 hours to efficiently generate divacancy centers[23]. In the PL6 divacancy magnetic detection experiments, we used a homemade confocal system combined with microwave and magnetic field systems. A 914 nm laser is used to polarize and read out the divacancy center spin state. A 980 nm dichroic mirror is applied to reflect the laser and an infrared objective (NA=0.65) is utilized to focus the laser on SiC. The PL6 fluorescence is collected by the same objective and transferred to a photoreceiver (Femto, OE-200-IN1) through an optical fiber. We use a 50 μm copper wire for microwave transmission to manipulate PL6 spin state. We use a standard lock-in measuring method to detect spin signal in the experiments[16,23,25,34]. A metal ceramic heater (HT24S, Thorlabs) and a platinum resistive temperature sensor (TH100PT, Thorlabs) are employed to control and calibrate SiC sample temperature[28].

**Acknowledgments**

This work was supported by National Key Research and Development Program of China (MOST) (Grant No. 2022YFA1405100) and National Natural Science Foundation of China (NSFC) (Grant No. 52172272, 11975221) and Science Specialty Program of Sichuan University (2020SCUNL210).

**Competing interests**